\documentclass[11pt,dvips]{article}
cm \voffset = -2truecm

\usepackage{graphicx}
\usepackage{amssymb}
\usepackage{latexsym}

\begin{document}
\begin{titlepage}
\setcounter{page}{1}
\renewcommand{\thefootnote}{\fnsymbol{footnote}}

\vspace{5mm}
\begin{center}

 {\Large \bf Bosonic and $k$-fermionic coherent states \\ 
             for a class of polynomial Weyl-Heisenberg algebras}
%Generalized Grassasmann coherent states for qudit systems

\vspace{1.5cm}

{\bf M. Daoud$^{1,2,3,4}$ and M. R. Kibler$^{1,2,3}$}

\vspace{0.8cm}

\noindent$^1$ Universit\'e de Lyon, 69361 Lyon, France \\
$^2$ Universit\'e Claude Bernard Lyon 1, 69622 Villeurbanne, France \\
$^3$ CNRS/IN2P3, Institut de Physique Nucl\'eaire, 69622 Villeurbanne, France \\
$^4$ D\'epartement de Physique, Facult\'e des Sciences, Agadir,
Morocco

\vspace{0.5cm} \noindent E-mail : m\_daoud@hotmail.com and
m.kibler@ipnl.in2p3.fr

\vspace{1.5cm}

\begin{abstract}

The aim of this article is to construct {\it \`a la} Perelomov and {\it \`a la} Barut--Girardello 
coherent states for a polynomial Weyl-Heisenberg algebra. This generalized Weyl-Heisenberg algebra, 
noted ${\cal A}_{ \{ \kappa \} }$, depends on $r$ real parameters and is an extension of the 
${\cal A}_{ \kappa }$ one-parameter algebra (Daoud M and Kibler M R 2010 {\em J. Phys. A: Math. Theor.} 
\textbf{43} 115303) which covers the cases of the $su(1,1)$ algebra (for $\kappa > 0$), the $su(2)$ 
algebra (for $\kappa < 0$) and the $h_4$ ordinary Weyl-Heisenberg algebra (for $\kappa = 0$). For 
finite-dimensional representations of ${\cal A}_{ \{ \kappa \} }$ and ${\cal A}_{ \{ \kappa \} , s }$, 
where ${\cal A}_{ \{ \kappa \} , s }$ is a truncation of order $s$ of ${\cal A}_{ \{ \kappa \} }$ in 
the sense of Pegg--Barnett, a connection is established with $k$-fermionic algebras (or quon 
algebras). This connection makes it possible to use generalized Grassmann variables for constructing 
certain coherent states. Coherent states of the Perelomov type are derived for 
infinite-dimensional representations of ${\cal A}_{ \{ \kappa \} }$ and for finite-dimensional  
representations of ${\cal A}_{ \{ \kappa \} }$ and ${\cal A}_{ \{ \kappa \} , s}$ through a Fock--Bargmann 
analytical approach based on the use of complex (or bosonic) variables. The same approach is applied for 
deriving coherent states of the Barut--Girardello type in the case of infinite-dimensional representations 
of ${\cal A}_{ \{ \kappa \} }$. In contrast, the construction of {\it \`a la} Barut--Girardello coherent 
states for finite-dimensional representations of ${\cal A}_{ \{ \kappa \} }$ and ${\cal A}_{ \{ \kappa \} , s }$
can be achieved solely at the price to replace complex variables by generalized Grassmann (or $k$-fermionic) 
variables. Some of the results are applied to $su(2)$, $su(1,1)$ and the harmonic oscillator (in a truncated or 
not truncated form).

\end{abstract}
\end{center}
\end{titlepage}

\newpage

\section{Introduction}

The coherent states play an essential role in many areas of physics 
(see \cite{Ali1, Gazeau2, Gazeau1, Klauder1, Perelomov1, Sudarshan}). They
were initially introduced by Schr\"odinger as quantum states whose 
dynamics is similar to a classical one \cite{Shrodinger}. There are 
three approaches to define coherent states for a quantum system
described by a (Lie) group or algebra (see for instance \cite{Zhang}). The 
Perelomov approach defines coherent states by the action of group 
elements on a fiducial state of the representation space of a group 
\cite{Perelomov2, Perelomov1} (see also \cite{Gilmore}). The
Barut--Girardello approach  defines coherent states as
eigenstates of a lowering group generator \cite{Barut}. The third
approach is based on the minimization of Robertson-Schr\"odinger
uncertainty relations for Hermitian generators of a group
\cite{Shrodinger1, Robertson} and leads to the so-called
intelligent states \cite{Aragone2, Aragone1}. The three approaches 
generally yields different coherent states except for the harmonic 
oscillator for which they give equivalent results (the so-called 
Glauber states \cite{Glauber}).

In this paper, we shall be concerned with the construction of Perelomov and Barut--Girardello coherent states 
for a polynomial Weyl-Heisenberg algebra.
%%%, noted ${\cal A}_{ \{ \kappa \} }$ in the following. 
It is known that the Perelomov approach can be used for any finite- or infinite-dimensional representation space of the 
considered group or algebra while the Barut--Girardello approach applies only to infinite-dimensional representation 
space \cite{Barut, Perelomov2, Perelomov1} (see also \cite{Fujii}). It was one of the object of this work to find a 
way to define Barut--Girardello coherent states for a finite-dimensional representation space. 

The original strategy adopted here is as follows. For Perelomov coherent states, we work with a Fock--Bargmann space 
for which the annihilation operator of the polynomial Weyl-Heisenberg algebra acts as derivative by a complex variable 
both in the finite- and infinite-dimensional cases. For Barut--Girardello coherent states, the creation operator acts 
on the Fock--Bargmann space as multiplication by a complex variable in the infinite-dimensional case and as 
multiplication by a generalized Grassmann variable in the finite-dimensional case. 

The paper is organized as follows. Section 2 deals with a polynomial Weyl-Heisenberg algebra, a $r$-parameter algebra 
denoted ${\cal A}_{ \{ \kappa \} }$ with $\{ \kappa \} \equiv \{ \kappa_1, \kappa_2, \cdots, \kappa_r \}$, which generalizes 
the ${\cal A}_{ \kappa }$ one-parameter oscillator algebra worked out in \cite{Atakishiyev, daoud-kibler2}. Such an algebra 
is shown to have finite- or infinite-dimensional representations depending on the values of the 
$\{ \kappa_1, \kappa_2, \cdots, \kappa_r \}$ parameters. In the case of an infinite-dimensional representation, it is shown how to restrict  
${\cal A}_{ \{ \kappa \} }$ to a ${\cal A}_{ \{ \kappa \} , s}$ truncated algebra with a representation of finite dimension 
$s$. In order to define Barut--Girardello coherent states in the finite-dimensional case, Section 3 is devoted to a review  
of $k$-fermion operators (which generalize ordinary fermion operators) and their realization in terms of generalized Grassmann 
variables. A connection between a $k$-fermionic algebra and either ${\cal A}_{ \{ \kappa \} }$ with a finite-dimensional 
representation or ${\cal A}_{ \{ \kappa \} , s}$ is studied in Section 4. Section 5 and 6 deal with the derivation of Perelomov 
and Barut--Girardello coherent states for ${\cal A}_{ \{ \kappa \} }$ and ${\cal A}_{ \{ \kappa \} , s}$. At this point, a word 
of explanation of the title is in order. The Perelomov states derived in the present paper in finite and infinite dimensions are 
of bosonic type (i.e., labeled by a complex variable) while the Barut-Girardello states are of $k$-fermionic type (i.e. labeled 
by a Grassmann variable) in finite dimension and of bosonic type in infinite dimension.  

%%%%%%%%%%%%%%%%%%%%%%%%%%%%%%%%%%%%%%%%%%%%%%%%%%%%%%%%%%%%%%%%%%%%%%%%%
\section{Generalized Weyl-Heisenberg algebra}
%%%%%%%%%%%%%%%%%%%%%%%%%%%%%%%%%%%%%%%%%%%%%%%%%%%%%%%%%%%%%%%%%%%%%%%%%

In the last two decades, due to the development of the theory of quantum algebras, 
non-linear extensions of the usual Weyl-Heisenberg algebra (or oscillator algebra) 
attracted a lot of attention. Many variants of generalized Weyl-Heisenberg 
algebras were proposed and studied from different viewpoints (e.g., see 
\cite{Daskaloyannis, Katriel, Quesne}). All generalized oscillators can be unified 
into a common framework by taking the product of the creation and annihilation operators 
(which is the $N$ number operator for the usual oscillator) to be $F(N)$,  
where the $F$ structure function characterizes the generalization scheme. 
Different structure functions correspond to different deformed oscillators. Of 
particular interest are the so-called polynomial Weyl-Heisenberg algebras in which 
$F(N)$ is a polynomial in $N$ and consequently the commutator of the annihilation 
and creation operators becomes a polynomial in $N$ too. From a physical point of 
view, such polynomial Weyl-Heisenberg algebras offer the advantage of dealing with
quantum systems with non-linear discrete spectrum \cite{Carballo, daoud-kibler2, El Kinani2, Fernandez}.

%%%%%%%%%%%%%%%%%%%%%%%%%%%%%%%%%%%%%%%%%%%%%%%%%%%%%%%%%%%%%%%%%%%%%%%%%
\subsection{The ${\cal A}_{\{\kappa\}}$ algebra}
%%%%%%%%%%%%%%%%%%%%%%%%%%%%%%%%%%%%%%%%%%%%%%%%%%%%%%%%%%%%%%%%%%%%%%%%%

We start with the generalized Weyl-Heisenberg algebra on $\mathbb{C}$ spanned by 
the linear operators $a^-$ (annihilation operator), $a^+$ (creation operator) 
and $N$ (number operator) satisfying the commutation relations 
	\begin{eqnarray}
	[a^- , a^+] =  G(N) \qquad 
	[N, a^-] = - a^- \qquad 
	[N, a^+] = + a^+ 	
	\label{algebre}
	\end{eqnarray}
and the Hermitian conjugation conditions
	\begin{eqnarray}
	a^+ = (a^-)^{\dagger} \qquad N = N^{\dagger}.
	\label{hermiticity conditions}
	\end{eqnarray}
The $G$ function in (\ref{algebre}) is such that 
	\begin{eqnarray}
	G(N) = \big(G(N)\big)^{\dagger}.
	\label{hermiticity of G}
	\end{eqnarray}
Of course, the case $G(N) = I$, where $I$ is the identity operator, corresponds 
to the usual Weyl-Heisenberg algebra or harmonic oscillator algebra. Various realizations 
of $G$ are known in the literature
\cite{Carballo, newdaokib, daoud-kibler1, daoud-kibler2, 
Daskaloyannis, Fernandez, Gavrilik, Katriel, Man'ko, Quesne}. In 
the present paper, we shall be concerned with a class of polynomial Weyl-Heisenberg algebras 
characterized by   
	\begin{eqnarray}
	G(N) = F(N+I) - F(N)
	\label{G as difference of two F}
	\end{eqnarray}
with the $F$ function defined by 
	\begin{eqnarray}
	F(N) = N [ I + \kappa_1(N-I)] 
	         [ I + \kappa_2(N-I)] \cdots 
	         [ I + \kappa_r(N-I)] 
	\label{generalformofF}
	\end{eqnarray}
where the $\kappa_i$'s ($i = 1, 2, \cdots , r$) are real parameters (for instance, see \cite{Gavrilik}). We note 
${\cal A}_{\{\kappa\}}$, with $\{\kappa\} \equiv \{\kappa_1, \kappa_2, \cdots, \kappa_r\}$, the 
generalized Weyl-Heisenberg algebra (or generalized oscillator algebra) defined via 
(\ref{algebre})-(\ref{generalformofF}). 

The $F(N)$ polynomial of order $r+1$ with respect to $N$ can be developed as 
	\begin{eqnarray}
	F(N) = N \sum_{i=0}^{r} s_i (N-I)^i 
	\label{development of F(N)}
	\end{eqnarray}
in terms of the coefficients (totally symmetric under permutation group $S_r$) 
	\begin{eqnarray}
	s_0 = 1 \qquad s_i = \sum_{j_1 < j_2 < \cdots < j_i}  \kappa_{j_1} \kappa_{j_2} \cdots \kappa_{j_i} \quad 
	(i = 1, 2, \cdots , r)
	\label{vieta formula}
	\end{eqnarray}
where the indices $j_1, j_2, \cdots , j_i$ take the values $1, 2, \cdots , r$. Then, the $G(N)$ operator can be written 
	\begin{eqnarray}
	G(N) = I + \sum_{i = 1}^r s_i \bigg[ (N+I)N^i - N(N-I)^i \bigg]
	\label{development of G(N)}
	\end{eqnarray}
which clearly indicates that ${\cal A}_{\{\kappa\}}$ with $\{\kappa\} \equiv \{0, 0, \cdots, 0\}$	coincides with 
the usual Weyl-Heisenberg algebra. 

The ${\cal A}_{\{\kappa\}}$ $r$-parameter algebra covers the cases of (i) the extended 
harmonic oscillator algebra \cite{Quesne}, (ii) the fractional oscillator 
algebra \cite{newdaokib}, and (iii) the $W_k$ algebra introduced in the 
context of fractional supersymmetric quantum mechanics of order $k$ 
\cite{daoud-kibler1, Kibler}. As a particular case, algebra ${\cal A}_{\{\kappa\}}$ 
with $\kappa_1 = \kappa$ and $r = 1$ is nothing but the ${\cal A}_{\kappa}$ 
algebra worked out in \cite{daoud-kibler2} and corresponding to 
	\begin{eqnarray}
	G(N) = I + 2\kappa N. 
	\label{G(N) de Akappa}
	\end{eqnarray}
Algebra ${\cal A}_{\kappa}$ defined by (\ref{algebre}), (\ref{hermiticity conditions}) and (\ref{G(N) de Akappa}) turns 
out to be of particular interest when dealing with dynamical symmetries of some exactly solvable quantum systems. More 
precisely, ${\cal A}_{\kappa = 0}$ corresponds to the usual oscillator system while ${\cal A}_{\kappa < 0}$ and 
${\cal A}_{\kappa > 0}$ are relevant to the Morse and P\"oschl-Teller systems, respectively 
\cite{daoud-kibler1, daoud-kibler2}. Note also that the ${\cal A}_{\kappa}$ one-parameter algebra provides a unified 
scheme to deal with the $su(2)$ algebra (for $\kappa < 0$), the $su(1,1)$ algebra (for $\kappa > 0$), and the usual 
Weyl-Heisenberg algebra (for $\kappa = 0$) \cite{Atakishiyev, daoud-kibler2}. More generally, the ${\cal A}_{ \{ \kappa \} }$ 
algebra can be viewed as a special class of the polynomial extensions of $su(2)$ and $su(1,1)$ discussed in \cite{Bonatsos} 
and \cite{Shreecharan}, respectively.

%%%%%%%%%%%%%%%%%%%%%%%%%%%%%%%%%%%%%%%%%%%%%%%%%%%%%%%%%%%%%%%%%%%%%%%%%%
\subsection{The ${\cal F}_{\{ \kappa \}}$ representation space}
%%%%%%%%%%%%%%%%%%%%%%%%%%%%%%%%%%%%%%%%%%%%%%%%%%%%%%%%%%%%%%%%%%%%%%%%%%

Let ${\cal F}_{\{\kappa\}}$ be the Fock-Hilbert space on which operators 
$a^-$, $a^+$ and $N$ act. This space is spanned by a complete set 
$\{ \vert n \rangle \ : \ n = 0, 1, 2, \cdots \}$ of orthonormal states 
which are eigenstates of operator $N$. It can be formally verified that 
the relations 
	\begin{eqnarray}
	&& a^-\vert n \rangle =\sqrt{F(n)}   e^{{+i [F(n) - F(n-1)]\varphi}} \vert n - 1 \rangle 
	\label{action sur les n amoins} \\ 
	&& a^+\vert n \rangle =\sqrt{F(n+1)} e^{{-i [F(n+1) - F(n)]\varphi}} \vert n + 1 \rangle 
	\label{action sur les n} \\
	&& a^-\vert 0 \rangle =0 \qquad N  \vert n \rangle = n \vert n \rangle
	\label{action sur les n amoins et N}
	\end{eqnarray}
define an Hilbertian representation of ${\cal A}_{\{\kappa\}}$. In Eqs.~(\ref{action sur les n amoins}) and (\ref{action sur les n}), $\varphi$ is an arbitrary real parameter and the $F$ structure function satisfies 
the conditions 
	\begin{eqnarray}
	F(0) = 0 \qquad F(n) > 0 \quad  (n=1, 2, 3, \cdots)
	\label{positivity}
	\end{eqnarray}
and 
	\begin{eqnarray}
	F(n+1) - F(n) = G(n) \iff F(n) = \sum_{m=0}^{n-1} G(m). 
	\end{eqnarray}
The dimension of the ${\cal F}_{\{\kappa\}}$ space can be finite or infinite. It is determined 
by condition (\ref{positivity}) which in turn depends on parameters $\kappa_i$ ($i = 1, 2, \cdots, r$). 
In the rest of this paper, we shall assume that  
	\begin{eqnarray}
	\kappa_1 \in \mathbb{R} \qquad \kappa_i \in \mathbb{R}_+  \quad (i = 2, 3, \cdots, r).   
	\label{signes de kappa}
	\end{eqnarray}
Hence, we have  	
	\begin{eqnarray}
	F(n) > 0 \ \Rightarrow \ 1 + \kappa_1(n-1) > 0 \quad (n \not= 0)
	\label{condition}
	\end{eqnarray}	
so that the dimension of the considered representation of ${\cal A}_{\{\kappa\}}$ is completely fixed by 
the sign of $\kappa_1$ in the following way. 
\begin{itemize}
	\item For $\kappa_1 \geq 0$, the ${\cal F}_{\{\kappa\}}$ space is infinite-dimensional. In this case, we have 
	\begin{eqnarray}
	F(n) = n \prod_{i = 1}^{r} [1 + \kappa_i(n-1)]
	\label{Fn infini} 
	\end{eqnarray}	
which gives $F(n) = n$ for the harmonic oscillator.
	\item For $\kappa_1   <  0$, there exists a finite number of states satisfying condition 
(\ref{condition}). In this case, $n$ can take the values 
	\begin{eqnarray}
	n = 0, 1, \cdots, E(-\frac{1}{\kappa_1}) 
	\end{eqnarray}
where we use $E(x)$ to denote the integer part of $x$. From now on, we shall assume that 
$-1/\kappa_1 \in \mathbb{N}^*$ when $\kappa_1 < 0$. Therefore, the dimension of 
${\cal F}_{\{\kappa\}}$ with $\kappa_1 < 0$ is 
	\begin{eqnarray}
	d = 1 - \frac{1}{\kappa_1} \qquad d \in \mathbb{N} \setminus \{ 0,1 \}. 
	\label{dimension}
	\end{eqnarray}
Note that Eq.~(\ref{dimension}) easily follows from ${\rm Tr} \> G(N) = 0$ that holds 
for $\kappa_1 < 0$ and $\kappa_i \geq 0$ ($i = 2, 3, \cdots, r$). In this case, we have 
	\begin{eqnarray}
	F(n) = n \frac{d-n}{d-1} \prod_{i = 2}^{r} [1 + \kappa_i(n-1)]
	\label{Fn fini} 
	\end{eqnarray} 
which gives $F(n) = n (d-n)/(d-1)$ for $r=1$ like in \cite{daoud-kibler2}.
\end{itemize}

The finiteness of the ${\cal F}_{\{\kappa\}}$ space  for $\kappa_1 < 0$ and 
$\kappa_i \geq 0$ ($i = 2, 3, \cdots, r$) induces properties of $a^-$ 
and $a^+$ which differ from those corresponding to an infinite-dimensional 
space. Along this line, we have the limit condition
	\begin{eqnarray}
	a^+ \vert d-1 \rangle = 0
	\end{eqnarray}
and the nilpotency relations
	\begin{eqnarray}
	(a^-)^{d} = (a^+)^{d} = 0 
	\label{nilpotency1}
	\end{eqnarray}
within the ${\cal F}_{\{\kappa\}}$ space generated by 
$\{ \vert 0 \rangle, \vert 1 \rangle, \cdots, \vert d-1 \rangle \}$.

%%%%%%%%%%%%%%%%%%%%%%%%%%%%%%%%%%%%%%%%%%%%%%%%%%%%%%%%%%%%%%%%%%%%%%%%%%%%%%%%%%%%%%%%%%
\subsection{The $H$ Hamiltonian}
%%%%%%%%%%%%%%%%%%%%%%%%%%%%%%%%%%%%%%%%%%%%%%%%%%%%%%%%%%%%%%%%%%%%%%%%%%%%%%%%%%%%%%%%%%

Going back to the general case where $\kappa_1 \in \mathbb{R}$ and 
$\kappa_i \in \mathbb{R}_+$ ($i = 2, 3, \cdots, r$), we have
	\begin{eqnarray}
	a^+a^- \vert n \rangle  = F(n) \vert n \rangle.
	\end{eqnarray}
We can thus introduce the operator 
	\begin{eqnarray}
	H \equiv F(N) = a^+a^-
	\end{eqnarray}
which generalizes the Hamiltonian for the one-dimensional harmonic oscillator 
up to an additive constant. We refer $H$ to as the Hamiltonian associated with 
generalized oscillator algebra ${\cal A}_{\{\kappa\}}$. Note that $F(N) = N$ 
for the usual harmonic oscillator which corresponds to $\kappa_i = 0$ 
($i = 1, 2, \cdots, r$). For $\kappa_1 \in \mathbb{R}^*$ and 
$\kappa_i = 0$ ($i = 2, 3, \cdots, r$), the $H$ Hamiltonian is quadratic 
in $N$ and can be related to some exactly solvable quantum systems as the 
Morse and P\"oschl-Teller systems \cite{daoud-kibler2}. For higher order in $N$, 
$H$ can be used to describe some non linear effects as for instance the isotopic effect in vibrational spectra of diatomic molecules (cf. \cite{Bergeron}). 
It is also relevant for dealing with higher-order supersymmetric harmonic oscillators 
\cite{Fernandez}. From a general point of view, it is interesting to note that energy gap $F(n+1)-F(n)$ between two consecutive states  behaves like $n^r$; therefore, for an infinite-dimensional space, we can ignore, in a perturbative scheme, the highly excited states with $n$ large and thus work with a subspace of finite dimension. 

%%%%%%%%%%%%%%%%%%%%%%%%%%%%%%%%%%%%%%%%%%%%%%%%%%%%%%%%%%%%%%%%%%%%%%%%%%%%%%%%%%%%%%%%%%
\subsection{The ${\cal A}_{\{ \kappa \},s}$ truncated algebra}
%%%%%%%%%%%%%%%%%%%%%%%%%%%%%%%%%%%%%%%%%%%%%%%%%%%%%%%%%%%%%%%%%%%%%%%%%%%%%%%%%%%%%%%%%%

As discussed above, in the case where $\kappa_i \geq 0$ ($i = 1, 2, \cdots, r$), the 
${\cal F}_{\{ \kappa \}}$ representation space of the ${\cal A}_{\{ \kappa \}}$ algebra is 
infinite-dimensional. In this case, to restrict ${\cal A}_{\{ \kappa \}}$ to a {\it truncated} algebra 
acting on a subspace of ${\cal F}_{\{ \kappa \}}$, we introduce operators $a^-(s)$ and $a^+(s)$
	\begin{eqnarray}
 	&& a^+(s)= a^+   - \sum_{n=s}^{\infty} \sqrt{F(n)} e^{-i [F(n)-F(n-1)]\varphi}\vert n \rangle \langle n-1 \vert  
 	\label{operators tronques aplus} \\
 	&& a^-(s)= a^-   - \sum_{n=s}^{\infty} \sqrt{F(n)} e^{+i [F(n)-F(n-1)]\varphi}\vert n-1 \rangle \langle n \vert
	\label{operators tronques}
	\end{eqnarray}
(we follow here the usual tradition in speaking of truncated algebra but it should be realized that the 
truncation applies to the representation space). They satisfy the following commutation relations
	\begin{eqnarray}
	[a^-(s) , a^+(s)] = G_s(N) - F(s) \vert s-1 \rangle \langle s-1 \vert \qquad 
	[N , a^{\pm}(s)] = \pm a^{\pm}(s)
	\label{thetruncatedalgebra}
	\end{eqnarray}
where
	\begin{eqnarray}
	G_s(N) = \sum_{n=0}^{s-1} [F(n+1)-F(n)] \vert n \rangle \langle n \vert
	\end{eqnarray}
and thus span with $N$ an algebra that we denote ${\cal A}_{\{ \kappa \}, s}$. The ${\cal A}_{\{ \kappa \}, s}$ 
algebra is refered to as truncated generalized oscillator algebra of order $s$. It generalizes algebra 
${\cal A}_{\kappa , s}$ defined in \cite{daoud-kibler2} which admits as a particular case the truncated 
Weyl-Heisenberg algebra introduced by Pegg end Barnett \cite{Pegg}. Indeed, the ${\cal A}_{\kappa , s}$ 
algebra with $\kappa = 0$ is identical to the Pegg--Barnett oscillator algebra.

It is easy to check that
	\begin{eqnarray}
	& & a^-(s) \vert n \rangle = \sqrt{F(n)}   e^{{+i [F(n) - F(n-1)] \varphi }} \vert n-1 \rangle 
	\label{action des b sur les n bmoins} \\
	& & a^+(s) \vert n \rangle = \sqrt{F(n+1)} e^{{-i [F(n+1) - F(n)] \varphi }} \vert n+1 \rangle 
	\label{action des b sur les n} \\
	& & a^-(s) \vert 0 \rangle = 0 \qquad a^+(s) \vert s-1 \rangle = 0 \qquad N \vert n \rangle = n \vert n \rangle 
	\label{action des b sur les n bmoins et N} 
	\end{eqnarray}
for $n = 0, 1, \cdots, s-1$. It follows that Eqs.~(\ref{action des b sur les n bmoins})-(\ref{action des b sur les n bmoins et N}) 
define a representation of ${\cal A}_{\{ \kappa \} , s}$  on the ${\cal F}_{\{ \kappa \} , s}$ space generated by the 
$\{ \vert 0 \rangle, \vert 1 \rangle, \cdots, \vert s-1 \rangle \}$ set. In this $s$-dimensional representation, operators 
$a^-(s)$ and $a^+(s)$ satisfy 
	\begin{eqnarray}
	[a^-(s)]^s = [a^+(s)]^s = 0
	\label{nilpotency2}
	\end{eqnarray}
to be compared with (\ref{nilpotency1}). It should be noted that Eqs.~(\ref{nilpotency1}) and (\ref{nilpotency2}) 
are identical for $d = s = k \in \mathbb{N} \setminus \{ 0,1 \}$ to the nilpotency relations describing the so-called 
$k$-fermions that are objects interpolating between fermions (for $k = 2$) and bosons (for $k \to \infty$) \cite{DHKrusse, Daoud}. 

%%%%%%%%%%%%%%%%%%%%%%%%%%%%%%%%%%%%%%%%%%%%%%%%%%%%%%%%%%%%%%
\section{Introducing $k$-fermions}
%%%%%%%%%%%%%%%%%%%%%%%%%%%%%%%%%%%%%%%%%%%%%%%%%%%%%%%%%%%%%%

%%%%%%%%%%%%%%%%%%%%%%%%%%%%%%%%%%%%%%%%%%%%%%%%%%%%%%%%%%%%%%
\subsection{The $k$-fermionic algebra}
%%%%%%%%%%%%%%%%%%%%%%%%%%%%%%%%%%%%%%%%%%%%%%%%%%%%%%%%%%%%%%

We first define a quon algebra. The $A_q$ quon algebra is generated by 
an annihilation operator ($f_-$), a creation operator ($f_+$) and a number 
operator ($N$) with the relations \cite{DHKrusse, Daoud, KibSIGMA}
	\begin{eqnarray}
	f_- f_+ - q f_+ f_- = I \qquad 
	[N , f_-] = - f_- \qquad 
	[N , f_+] = + f_+ \qquad 
	N = N^{\dagger}
	\label{commut des oper f}
	\end{eqnarray}
and 
	\begin{eqnarray}
	(f_-)^k = (f_+)^k = 0 
	\label{nilpotencyf}
	\end{eqnarray}
where $q$ is primitive $k$th root of unity: 
	\begin{eqnarray}
	q = {\rm exp} \left( \frac{2 \pi {i}}{k} \right) \qquad k \in \mathbb{N} \setminus \{ 0,1 \}. 
	\end{eqnarray}
Note that the $q$-commutation 
relation in Eq.~(\ref{commut des oper f}) is satisfied by
	\begin{eqnarray}
	f_- f_+ = \left[ N + I \right]_q  \qquad  f_+ f_- = \left[ N \right]_q
	\label{deformed N and Nplus1} 
	\end{eqnarray}
with
	\begin{eqnarray}
	[X]_q = {1 - q^X \over 1 - q}
	\label{xdeformed}
	\end{eqnarray}
where $X$ may be an operator or a complex number.

It is important to realize that quon algebra $A_q$ differs from that introduced by Arik and Coon \cite{Arik}, 
Biedenharn \cite{Biedenharn} and Macfarlane \cite{Macfarlane}
because $q$ is not here a real number and operators $f_-$ and $f_+$ satisfy nilpotency relations (of order 
$k$). Obviously, Eq.~(\ref{commut des oper f}) shows that $f_-$ and $f_+$ cannot be connected by Hermitian 
conjugation (i.e., $f_+ \not= (f_-)^{\dagger}$) when $q$ is a root of unity except for $k=2$ (i.e., $q = -1$) or 
$k \to \infty$ (i.e., $q = +1$). The $A_{-1}$ algebra corresponds to ordinary fermion operators with 
$(f_+)^2 = (f_-)^2 = 0$, a relation that reflects the Pauli exclusion principle. In the limiting case where 
$k \to \infty$, the $A_{+1}$ algebra corresponds to ordinary boson operators; in this case, Eq.~(\ref{commut des oper f}) 
describes the usual harmonic oscillator algebra. For $q$ a primitive $k$-th root of unity different from $\pm 1$, operators 
$f_-$ and $f_+$ are referred to as $k$-fermion operators.

It is a simple exercise to check that the actions 
	\begin{eqnarray}
&&f_- \vert n \rangle = \left( \left[ n   \right]_q \right)^{{1 \over2}} | n-1 \rangle \qquad f_- \vert 0   \rangle = 0 
	\label{action f moins} \\
&&f_+ \vert n \rangle = \left( \left[ n+1 \right]_q \right)^{{1 \over2}} | n+1 \rangle \qquad f_+ \vert k-1 \rangle = 0 
	\label{action f plus}
	\end{eqnarray}
and
	\begin{eqnarray}
	N \vert n \rangle = n \vert n \rangle
	\end{eqnarray}
on a Fock-Hilbert space of dimension $k$, noted $F_q$ and isomorphic with ${\cal F}_{ \{ \kappa \} }$ with $d = k$, 
defines a $k$-dimensional representation of $A_q$. Of course, $\dim F_{-1} = 2$ for fermions while $\dim F_{+1}$ is 
infinite for bosons. Any vector of $F_q$ can be obtained by repeated application of $f_+$ on the {\it ground} state: 
	\begin{eqnarray}
	\vert n \rangle = \frac{1}{ ([n]_q !)^{1 \over 2} } (f_+)^n \vert 0 \rangle \quad (n = 0, 1, \cdots, k-1)
	\end{eqnarray}
where the $[n]_q$-factorial is defined by
	\begin{eqnarray}
\lbrack 0 \rbrack_q ! = 1 \qquad 
\lbrack n \rbrack_q ! = \prod_{i=1}^n
\lbrack i \rbrack_q \quad n \in \mathbb{N} \setminus \lbrace 0 \rbrace 
	\end{eqnarray}
as is usual.

It was mentioned above that the relation $f_+ = (f_-)^{\dagger}$ does not hold in general. Thus, we introduce the operators 
	\begin{eqnarray}
 	f_+^+ = (f_+)^{\dagger} \qquad 
 	f_-^+ = (f_-)^{\dagger}. 
	\end{eqnarray}
The three operators $f_+^+$, $f_-^+$ and $N$ span the $A_{\bar q}$ quon algebra defined by 
	\begin{eqnarray}
	f_+^+ f_-^+ - {\bar q} f_-^+ f_+^+ = I \qquad 
	[N , f_+^+ ] = - f_+^+ \qquad 
	[N , f_-^+ ] = + f_-^+ \qquad 
	N = N^{\dagger}
	\label{commut des oper fconjugated}
	\end{eqnarray}
and
	\begin{eqnarray}
	(f_+^+)^k = (f_-^+)^k = 0. 
	\label{nilpotencyfconjugated}
	\end{eqnarray}
Operator $f_+^+$ plays the role of an annihilation operator and operator $f_-^+$ the
one of a creation operator for the $A_{\bar q}$ algebra. We can check that $f_+^+$ and
$f_-^+$ act on $F_q$ according to 
	\begin{eqnarray}
	&& f_+^+ \vert n \rangle = \left( \left[ n   \right]_{\bar q} \right)^{1 \over 2} | n-1 \rangle \qquad f_+^+ \vert 0   \rangle = 0 \\
	&& f_-^+ \vert n \rangle = \left( \left[ n+1 \right]_{\bar q} \right)^{1 \over 2} | n+1 \rangle \qquad f_-^+ \vert k-1 \rangle = 0.
	\end{eqnarray}
At this point, it is worth noticing that the relation
	\begin{eqnarray}
	f_- f_+^+ = q^{-{1 \over 2}} f_+^+ f_- \iff 
	f_+ f_-^+ = q^{+{1 \over 2}} f_-^+ f_+
	\label{mixed Aq Aqbar}
	\end{eqnarray}
holds on the $F_q$ space. We shall refer to $k$-fermionic algebra, noted $\Sigma_q$, the 
algebra generated by the ($f_+, f_-^+$) and ($f_-, f_+^+$) pairs together with the $N$ operator 
satisfying Eqs.~(\ref{commut des oper f}), 
(\ref{nilpotencyf}), 
(\ref{commut des oper fconjugated}), 
(\ref{nilpotencyfconjugated}) and
(\ref{mixed Aq Aqbar}).

%%%%%%%%%%%%%%%%%%%%%%%%%%%%%%%%%%%%%%%%%%%%%%%%%%%%%%%%%%%%%%%%%
\subsection{Grassmannian realization}
%%%%%%%%%%%%%%%%%%%%%%%%%%%%%%%%%%%%%%%%%%%%%%%%%%%%%%%%%%%%%%%%%

Equations (\ref{nilpotencyf}) and (\ref{nilpotencyfconjugated}) suggest that we look for a realization of 
the ($f_+, f_-^+$) and ($f_-, f_+^+$) pairs in terms of generalized Grassmann variables $(\theta, {\bar \theta})$ 
and their $q$- and ${\bar q}$-derivatives $(\partial_{\theta}, \partial_{\bar \theta})$. Generalized Grassmann 
variables $\theta$ and ${\bar \theta}$ of order $k$, introduced in connection with quantum groups and fractional 
supersymmetry, satisfy 
	\begin{eqnarray}
	\theta^k = {\bar \theta}^k = 0 
	\label{idempotence de teta}
	\end{eqnarray}
(see \cite{Ahn, de Azcarraga, Filippov, Le Clair, Majid, Rubakov}). The sets 
$\{ I, \theta , \cdots, \theta ^{k-1} \}$ and 
$\{ I, {\bar \theta}, \cdots, {\bar \theta} ^{k-1} \}$ span isomorphic Grassmann algebras. The 
$q$- and ${\bar q}$-derivatives are formally defined by 
	\begin{eqnarray}
	\partial_{\theta} f(\theta) = {f(q \theta) - f(\theta) \over (q - 1) \theta} \qquad
	\partial_{\bar \theta} g({\bar \theta}) = 
	{g({\bar q} {\bar \theta}) - g({\bar \theta}) \over ({\bar q} - 1) {\bar \theta}}
	\end{eqnarray}
where $f :       \theta  \mapsto f(      \theta )$ and 
      $g : {\bar \theta} \mapsto g({\bar \theta})$ are arbitrary functions. The 
$\partial_{\theta}$ and $\partial_{\bar \theta}$ operators satisfy 
	\begin{eqnarray}
	\partial_{\theta} \theta^n = [n]_q \> \theta^{n-1} \qquad
	\partial_{\bar \theta} {\bar \theta}^n = [n]_{\bar q} \> {\bar \theta}^{n-1}
	\end{eqnarray}
for $n = 0, 1, \cdots, k-1$. Hence, for functions $f$ and $g$ such that 
	\begin{eqnarray}
	f(\theta) = \sum_{n=0}^{k-1} a_n \theta^n \qquad
	g({\bar \theta}) = \sum_{n=0}^{k-1} b_n {\bar \theta}^n
	\end{eqnarray}
where the $a_n$ and $b_n$ coefficients in the expansions are complex numbers,
we easily show that 
	\begin{eqnarray}
	(\partial_{\theta})^k f(\theta) = 
	(\partial_{\bar \theta})^k g({\bar \theta}) = 0.
	\end{eqnarray}
As a consequence, we assume that the conditions 
	\begin{eqnarray}
	(\partial_{\theta})^k = 
	(\partial_{\bar \theta})^k = 0
	\end{eqnarray}
hold in addition to (\ref{idempotence de teta}). We can also define an integration process. Following 
Majid and Rodr\'\i guez-Plaza \cite{Majid}, we take
	\begin{eqnarray}
	\int {\theta} ^n  d{\theta} = 
	\int {\bar {\theta}}^n d{\bar {\theta}} = 0 \quad (n = 0, 1, \cdots, k-2) 
	\label{majid-formule1}
	\end{eqnarray}
and
	\begin{eqnarray}
	\int {\theta}^{k-1} d{\theta} = 
	\int {\bar {\theta}}^{k-1} d{\bar {\theta}} = 1 
	\label{majid-formule2}
	\end{eqnarray}
which gives the Berezin integration for the $k=2$ particular case ($k=2$ corresponds to the ordinary Grassmann 
variables used in supersymmetry).

As a result, we can show that the correspondences 
	\begin{eqnarray}
	f_-   \longrightarrow \partial_{\theta} \qquad 
	f_+   \longrightarrow           \theta  \qquad
	f_+^+ \longrightarrow \partial_{\bar \theta} \qquad 
	f_-^+ \longrightarrow {\bar          \theta}
	\label{Grassmann realization}
	\end{eqnarray}
provide us with a realization of the $q$- and $\bar q$-commutators in (\ref{commut des oper f}) and (\ref{commut des oper fconjugated})
and of the nilpotency relations in (\ref{nilpotencyf}) and (\ref{nilpotencyfconjugated}). Note that (\ref{mixed Aq Aqbar}) becomes 
	\begin{eqnarray}
	\theta {\bar \theta} = q^{+{1 \over 2}} {\bar \theta} \theta \qquad
	\partial_{\theta} \partial_{\bar \theta} = q^{-{1 \over 2}} \partial_{\bar \theta} \partial_{\theta}
	\label{eq56}
	\end{eqnarray}
in the realization afforded by (\ref{Grassmann realization}).

%%%%%%%%%%%%%%%%%%%%%%%%%%%%%%%%%%%%%%%%%%%%%%%%%%%%%%%%%%%%%%%%%%%%%%%%%%%%%%%%%%%%
\section{From generalized Weyl-Heisenberg algebra to quon algebras}
%%%%%%%%%%%%%%%%%%%%%%%%%%%%%%%%%%%%%%%%%%%%%%%%%%%%%%%%%%%%%%%%%%%%%%%%%%%%%%%%%%%%

\subsection{Obtaining the $A_q$ algebra}

We may ask how to pass from the ${\cal A}_{ \{ \kappa \} }$ generalized Weyl-Heisenberg algebra to the 
$A_q$ quon algebra. Since $A_q$ admits a representation of dimension $k$ ($q = \exp(2 \pi i / k)$), we 
shall consider in this section algebra ${\cal A}_{ \{ \kappa \} }$ with $\{ \kappa \}$ given by 
	\begin{equation}
	\kappa_1 = - \frac{1}{k-1} < 0 \qquad \kappa_i \geq 0 \quad (i=2, 3, \cdots, r)
	\label{kappa cas fini avec k}
	\end{equation}
with $k \in \mathbb{N} \setminus \{ 0,1 \}$. Our aim is to find three operators, 
say $A_-$, $A_+$ and $N_A$, acting on ${\cal F}_{\{ \kappa \}}$ and expressed as functions of generators 
$a_-$, $a_+$ and $N$ of ${\cal A}_{ \{ \kappa \} }$, such that $A_-$, $A_+$ and $N_A$ span $A_q$. For this 
purpose, we put 
	\begin{eqnarray}
	A_- = \sum_{i=1}^{k-1} C_i (a^+)^{i-1} (a^-)^{i} \qquad A_+ = a^+ \qquad N_A = N
	\label{definition-of-f}
	\end{eqnarray}
where the $C_i$'s are expansion coefficients to be determined. 

As a first step, in order to determine the $C_i$ coefficients, it is 
sufficient to require that 
 	\begin{eqnarray}
	A_+A_- = [N]_q
	\label{Aplus x Amoins}
	\end{eqnarray}
on the ${\cal F}_{\{ \kappa \}}$ representation space (cf. Eq.~(\ref{deformed N and Nplus1})). This yields 
 	\begin{eqnarray}
	\sum_{i=1}^{k-1} C_i (a^+)^{i} (a^-)^{i} \vert n \rangle = [n]_q \vert n \rangle.
	\label{action de AplusAmoins}
	\end{eqnarray}
A simple development of the lhs of (\ref{action de AplusAmoins}) leads to the 
following sytem of $k-1$ unknowns with $k-1$ equations: 
	\begin{eqnarray}
	\sum_{i=1}^{n} C_i \frac{F(n)!}{F(n-i)!} = [n]_q \quad (n = 1, 2, \cdots, k-1)
	\label{system}
	\end{eqnarray}
where the $F(n)$-factorial is defined by 
	\begin{equation}
	F(0)! = 1 \qquad F(n)! = \prod_{i=1}^n F(i) \quad n \in \mathbb{N} \setminus \{ 0 \}. 
	\label{F(n) factorial}
	\end{equation}
The $F(i)$'s occurring in (\ref{system}) are taken from (\ref{Fn fini}) with $d = k$. System (\ref{system}) 
can be written as 
	\begin{eqnarray}
T \left(
\begin{array}{c}
  C_1 \\
  C_2 \\
  \vdots \\
  C_{k-1} \\
\end{array}
\right) = \left(
\begin{array}{c}
  1 \\
  {[2]}_q \\
  \vdots \\
  {[k-1]}_q \\
\end{array}
\right)
	\label{system in matrix form}
	\end{eqnarray}
where $(k-1) \times (k-1)$-matrix $T$ reads
	\begin{eqnarray}
T = \left(%
\begin{array}{ccccc}
  F(1)! & 0 & 0 & \cdots & 0 \\
  F(2) & F(2)! & 0 & \cdots & 0 \\
  F(3)& F(3)F(2) & F(3)! & \cdots & 0 \\
  \vdots & \vdots & \vdots  & \vdots  & \vdots \\
  F(k-1)& F(k-1)F(k-2) & F(k-1)F(k-2)F(k-3) & \cdots & F(k)! \\
\end{array}%
\right)
	\end{eqnarray}	
Clearly 
	\begin{eqnarray}
	{\rm det} \> T =  F(1)! F(2)!\cdots F(k-1)! \not= 0
	\end{eqnarray}
so that there is a unique solution for the $C_i$ coefficients which can be calculated from
	\begin{eqnarray}
	\left(
	\begin{array}{c}
  C_1 \\
  C_2 \\
  \vdots \\
  C_{k-1} \\
	\end{array}
	\right) = T^{-1} 
	\left(
	\begin{array}{c}
  1 \\
  {[2]}_q \\
  \vdots \\
  {[k-1]}_q \\
	\end{array}
	\right)
	\label{inverse system}
	\end{eqnarray}
by inverting (\ref{system in matrix form}). 

As a second step, we prove that 
	\begin{eqnarray}
	A_-A_+ = [N + I]_q
	\label{Amoins x Aplus}
	\end{eqnarray}
holds for the operators $A_-$ and $A_+$ given by (\ref{definition-of-f}) with the $C_i$ coefficients 
derived from (\ref{system}). This can be seen as follows. We have 
	\begin{eqnarray}
	(a^+)^{i-1} (a^-)^{i} a^+ \vert n \rangle = 
	\cases{
	F(n+1)\frac{F(n)!}{F(n+1-i)!} \vert n \rangle \quad {\rm if} \quad i \leq n+1 \cr \cr 
  0 \quad {\rm if} \quad i > n+1. \cr
  }
	\end{eqnarray}
This implies
	\begin{eqnarray}
	A_-A_+ \vert n \rangle = \sum_{i = 1}^{n+1} C_i
	\frac{F(n+1)!}{F(n+1-i)!} \vert n \rangle.
	\label{action de AmoinsAplus sur n}
	\end{eqnarray}
By using (\ref{system}), Eq.~(\ref{action de AmoinsAplus sur n}) can be rewritten as 
	\begin{eqnarray}
	A_-A_+ \vert n \rangle = [n+1]_q \vert n \rangle
	\end{eqnarray}
which shows that (\ref{Amoins x Aplus}) is satisfied on ${\cal F}_{\{ \kappa \}}$.

The two preceding steps leads to 
	\begin{eqnarray}
	A_- A_+ - q A_+ A_- = I.
	\label{q-commutateur en A}
	\end{eqnarray}
Furthermore, it is trivial to check that 	
	\begin{eqnarray}	
	[N , A_{\pm}] = \pm A_{\pm} \qquad (A_-)^k = (A_+)^k = 0. 	
	\label{com de N et A plus nilpotence}
	\end{eqnarray}	
As a result, Eqs.~(\ref{q-commutateur en A}) and (\ref{com de N et A plus nilpotence}) 
correspond to Eqs.~(\ref{commut des oper f}) and (\ref{nilpotencyf}). Therefore, 
$A_-$, $A_+$ and $N_A$ generate quon algebra $A_q$ with $q = \exp(2 \pi i / k)$. 

We close this section with the action of annihilation operator $A_-$ and creation operator $A_+$ 
on ${\cal F}_{\{ \kappa \}}$:
\begin{eqnarray}
& & A_- \vert n \rangle =   \frac{[n]_q}{\sqrt{F(n)}} e^{{+i [F(n) - F(n-1)] \varphi }} \vert n-1 \rangle 
\qquad A_- \vert 0 \rangle = 0
\label{seconde action des f1} \\
& & A_+ \vert n \rangle =  \sqrt{F(n+1)} e^{{-i [F(n+1)- F(n)] \varphi }} \vert n+1 \rangle 
\qquad A_+ \vert k-1 \rangle = 0.
\label{seconde action des f2}
\end{eqnarray}	
The derivation of (\ref{seconde action des f2}) is immediate while the proof of (\ref{seconde action des f1})  
requires the use of (\ref{system}). 

\subsection{Obtaining the $A_{\bar q}$ quon algebra} 

A similar connection between ${\cal A}_{ \{ \kappa \} }$ and $A_{\bar q}$ can be obtained. By defining 
	\begin{eqnarray}
	A_+^+ = a^- \qquad A_-^+ = \sum_{i=1}^{k-1} {\overline C_i} (a^+)^{i} (a^-)^{i-1} \qquad N_A = N
	\label{definition-of-fadjoint}
	\end{eqnarray}
it straightforwardly follows that $A_+^+$, $A_-^+$ and $N_A$ span quon algebra $A_{\bar q}$. The action 
of $A_+^+$ and $A_-^+$ on ${\cal F}_{\{ \kappa \}}$ is given by 
	\begin{eqnarray}
	& & A_+^+ \vert n \rangle = \sqrt{F(n)}  e^{{+i [F(n) - F(n-1)] \varphi }} \vert n-1 \rangle 
	\qquad A_+^+ \vert 0 \rangle = 0
	\label{seconde action des f1conj} \\
	& & A_-^+ \vert n \rangle = \frac{[n+1]_{\bar q}}{\sqrt{F(n+1)}}   e^{{-i [F(n+1)- F(n)] \varphi }} \vert n+1 \rangle 
	\qquad A_-^+  \vert k-1 \rangle = 0
	\label{seconde action des f2conj}
	\end{eqnarray}	
cf.~(\ref{seconde action des f1}) and (\ref{seconde action des f2}). 

\subsection{Passage formulas}

The main result of Sections 6.1 and 6.2 is that triplets 
$\{ A_-, A_+, N \}$ and $\{ A_+^+, A_-^+, N \}$ span the 
$A_q$ and $A_{\bar q}$ quon algebras, respectively. This 
does not mean that operators $A_{\pm}$ and $A_{\pm}^+$ can 
be identified with fermion operators $f_{\pm}$ and $f_{\pm}^+$, 
respectively. Indeed, it can be verified that the following passage 
formulas 
	\begin{eqnarray}
	A_- =  \bigg( \frac{[N+I]_q}{F(N+I)}\bigg)^{\frac{1}{2}}    e^{+iG(N)\varphi} f_- \qquad
	A_+ = f_+ \bigg( \frac{F(N+I)}{[N+I]_q}\bigg)^{\frac{1}{2}} e^{-iG(N)\varphi}
	\end{eqnarray}
and 
	\begin{eqnarray}
	A_+^+ = \bigg( \frac{F(N+I)}{[N+I]_{\bar q}}\bigg)^{\frac{1}{2}}       e^{+iG(N)\varphi} f_+^+ \qquad
	A_-^+ = f_-^+ \bigg( \frac{[N+I]_{\bar q}}{F(N+I)}\bigg)^{\frac{1}{2}} e^{-iG(N)\varphi}  
	\end{eqnarray}
are compatible with the action of operators $f_{\pm}$, $f_{\pm}^+$,
$A_{\pm}$ and $A_{\pm}^+$ on the ${\cal F}_{ \{ \kappa \} }$.

Finally, note that the analysis presented in Section 4 can be extended to truncated 
Weyl-Heisenberg algebra ${\cal A}_{ \{ \kappa \} , s }$ modulo some obvious changes. 

%%%%%%%%%%%%%%%%%%%%%%%%%%%%%%%%%%%%%%%%%%%%%%%%%%%%%%%%%%%%%%%%%%%%%%%%%%%%%%%%%%%%%%%
\section{{\it \`a la} Perelomov coherent states}
%%%%%%%%%%%%%%%%%%%%%%%%%%%%%%%%%%%%%%%%%%%%%%%%%%%%%%%%%%%%%%%%%%%%%%%%%%%%%%%%%%%%%%%%

%%%%%%%%%%%%%%%%%%%%%%%%%%%%%%%%%%%%%%%%%%%%%%%%%%%%%%%%%%%%%%%%%%%%%
\subsection{Infinite-dimensional Fock-Hilbert space}
%%%%%%%%%%%%%%%%%%%%%%%%%%%%%%%%%%%%%%%%%%%%%%%%%%%%%%%%%%%%%%%%%%%%%

For $\kappa_i \geq 0$ ($i = 1,2, \cdots, r$), let us look for $\varphi$-dependent 
coherent states associated with the ${\cal A}_{\{ \kappa \}}$ algebra in the form 
	\begin{equation}
	\vert z , \varphi \rangle =  \sum_{n = 0}^{\infty} {\overline{a_n}} z^n \vert n \rangle
	\end{equation}
where $z$ is a complex variable and the $a_n$ coefficients depend on $\varphi$. Inspired by 
Bargmann \cite{Bargmann}, we assume that the basis vectors of the ${\cal F}_{\{ \kappa \}}$ 
infinite-dimensional space are realized as follows 
	\begin{equation}
	\vert n \rangle \longrightarrow a_{n} z^{n} \equiv \langle \bar z , \varphi \vert n \rangle 
	\label{correspondance1}
	\end{equation}
and the $a^-$ annihilation operator acts on ${\cal F}_{\{ \kappa \}}$ as a derivation 
according to
	\begin{equation}
	a^-\longrightarrow \frac{d}{dz}.
	\label{correspondance2}
	\end{equation}
Then, the application of the correspondence rules (\ref{correspondance1}) and 
(\ref{correspondance2}) to the expression of $a^- \vert n \rangle$ given by 
(\ref{action sur les n amoins}) yields the following recursion relation
	\begin{equation}
	n a_{n}= \sqrt{F(n)}e^{+i[F(n)-F(n-1)]\varphi}a_{n-1}
	\end{equation}
which leads to 
	\begin{equation}
	a_{n}= a_0 \frac{\sqrt{F(n)!}}{n!}e^{+iF(n)\varphi}
	\end{equation}
where $F(n)!$ follows from (\ref{F(n) factorial}) and (\ref{Fn infini}). As a result, we get the normalized state vectors 
	\begin{equation}
	\vert z , \varphi \rangle = {\cal N}^{-1} \sum_{n=0}^{\infty} 
  \frac{\sqrt{F(n)!}}{n!} z^n e^{-iF(n)\varphi} \vert n \rangle
	\label{KP-CS-infinite}
	\end{equation}
where the ${\cal N}$ normalization factor is formally given by
	\begin{equation}
	|{\cal N}|^2 = \sum_{n = 0}^{\infty} \frac{F(n)!}{(n!)^2} \vert z \vert^{2n}
	\label{squarred norm} 
	\end{equation}
subject to convergence. It should be observed that the $\vert z , \varphi \rangle$ 
vectors given by (\ref{KP-CS-infinite}) and (\ref{squarred norm}) with $\kappa_i = 0$ 
($i = 1, 2, \cdots, r$) and $\varphi = 0$ are coherent states for the usual harmonic 
oscillator. The series in (\ref{squarred norm}) diverges for $r \geq 2$ when 
$\kappa_i \not= 0$ ($i = 1, 2, \cdots, r$). For $r=1$, it converges in the disk 
	\begin{equation}
{\cal D} = \{ z \in \mathbb{C} \ : \ \vert z \vert < \frac{1}{\sqrt{\kappa_1}} \}.
	\end{equation}
Therefore, only the $r=1$ case deserves to be considered here (in contrast with Section 6.1). 

For $r=1$, let us consider the special case 
	\begin{equation}
\kappa_1 \equiv \kappa = \frac{1}{\ell} \qquad k \in \mathbb{N}^*.
	\label{un sur kappaun entier}
	\end{equation}
In this case, 
Eqs.~(\ref{correspondance1}) and (\ref{correspondance2}) can be completed by
\begin{equation}
N   \longrightarrow z \frac{d}{dz} \qquad 
a^+ \longrightarrow z \bigg( 1 + \frac{1}{\ell} z \frac{d}{dz}\bigg).
\end{equation}
The $\vert z , \varphi \rangle$ state vectors for the corresponding 
${\cal A}_{\kappa}$ algebra follow from (\ref{KP-CS-infinite})-(\ref{un sur kappaun entier}). We 
obtain 
	\begin{equation}
	\vert z , \varphi \rangle = {\cal N}^{-1} \sum_{n = 0}^{\infty}
	\sqrt{ \frac{1}{n!} \frac{(\ell + n - 1)!}{\ell^n (\ell-1)!} } z^n e^{-iF(n)\varphi}
	\vert n \rangle 
	\label{Perelomov-CS-r=1}
	\end{equation}
with
	\begin{equation}
	\vert {\cal N} \vert^2 =  \sum_{n = 0}^{\infty} \frac{1}{n!} \frac{ (\ell + n - 1)!}{\ell^n (\ell-1)!} \vert z \vert^{2n} 
	= \bigg(1 - \frac{|z|^2}{\ell}\bigg)^{-\ell}.
	\label{norme1r=1}
	\end{equation}
The $d\mu$ measure, assumed to be isotropic (i.e., $\vert z \vert$-dependent), for which states
(\ref{Perelomov-CS-r=1}) satisfy the over-completeness relation
	\begin{equation}
	\int d\mu (\vert z \vert) \vert z , \varphi \rangle \langle z , \varphi \vert = \sum_{n=0}^{\infty} \vert n \rangle \langle n \vert
	\end{equation}
can be easily derived from a reasoning similar to that used for the $su(1,1)$ Perelomov coherent states
(see for instance \cite{Brif, Perelomov1}). This leads to
	\begin{equation}
	d\mu (\vert z \vert) = \frac{1}{\pi} \frac{\ell-1}{\ell} \bigg(1 - \frac{\vert z \vert^2}{\ell} \bigg)^{-2} d^2z. 
	\label{measureKPr=1}
	\end{equation}
There are two other important properties of states (\ref{Perelomov-CS-r=1}). First, they are temporally stable, i.e.
	\begin{equation}
	e^{-i H t} | z , \varphi \rangle = | z, \varphi + t \rangle
	\end{equation}
holds for any real value of $t$. Second, the $| z , \varphi \rangle$ vector 
in (\ref{Perelomov-CS-r=1}) can be written 
	\begin{equation}
	\vert z , \varphi \rangle = {\cal N}^{-1}\exp( z a^+) \vert 0 \rangle.
	\end{equation}
Therefore, the $| z , \varphi \rangle$ vectors corresponding to $r = 1$ and 
$\kappa_1 \equiv \kappa = 1/\ell$ with $\ell \in \mathbb{N}^*$ result from the 
action of a displacement operator on the vacuum and are thus coherent states 
in the Perelomov sense \cite{Klauder1, Perelomov1}.

%%%%%%%%%%%%%%%%%%%%%%%%%%%%%%%%%%%%%%%%%%%%%%%%%%%%%%%%%%%%%%%%%%%%%
\subsection{Truncated Fock-Hilbert space}
%%%%%%%%%%%%%%%%%%%%%%%%%%%%%%%%%%%%%%%%%%%%%%%%%%%%%%%%%%%%%%%%%%%%%

As already mentioned, the norm given by (\ref{squarred norm}) diverges for $r \geq 2$ when 
$\kappa_i \not= 0$ ($i = 1, 2, \cdots, r$). However, by restricting $n$ to take the values 
$0, 1, \cdots, s-1$ in (\ref{squarred norm}), the norm is defined. This amounts to replace 
${\cal A}_{\{ \kappa \}}$ by the ${\cal A}_{\{ \kappa \} ,s}$ truncated algebra and to use 
the correspondence
	\begin{equation}
	a^-(s) \longrightarrow \frac{d}{dz}
	\end{equation}
which, for $r \geq 1$ and $s$ finite, can be supplemented with
	\begin{equation}
	N      \longrightarrow z\frac{d}{dz}, \qquad
	a^+(s) \longrightarrow z \bigg( 1 + \kappa_1 z \frac{d}{dz}\bigg) 
	                         \bigg( 1 + \kappa_2 z \frac{d}{dz}\bigg) \cdots
	                         \bigg( 1 + \kappa_r z \frac{d}{dz}\bigg).
	\label{corres pr N et aplus}
	\end{equation}
Calculations similar to those developed in Section 5.1 lead to the coherent states 
	\begin{equation}
	\vert z , \varphi \rangle = {\cal N}^{-1} \sum_{n = 0}^{s-1}
	\frac{\sqrt{F(n)!}}{n!} z^n e^{-iF(n)\varphi} \vert n \rangle
	\label{KP-CS-truncated}
	\end{equation}
with 
	\begin{equation}
	\vert {\cal N} \vert^2 = \sum_{n = 0}^{s-1} \frac{F(n)!}{(n!)^2} \vert z \vert^{2n}
	\label{norme1s}
	\end{equation}
where $F(n)!$ follows from (\ref{F(n) factorial}) and (\ref{Fn infini}). The states given by 
(\ref{KP-CS-truncated}) and (\ref{norme1s}) are coherent states in the Perelomov sence since 
	\begin{equation}
	\vert z , \varphi \rangle = {\cal N}^{-1} \exp [ z a^+(s) ] \vert 0 \rangle.
	\end{equation}
In addition, they are stable under time evolution.

The situation where 
	\begin{equation}
	\ell_i = \frac{1}{\kappa_i} \in \mathbb{N}^* \quad (i = 1, 2, \cdots, r)
	\label{un sur kappai entier}
	\end{equation}	
(each $\ell_i$ is assumed here to be a strictly positive integer) is especially interesting. In this 
case, $F(n)$ reads
	\begin{equation}
	F(n) = \frac{1}{\ell_1\ell_2\cdots \ell_r} n (\ell_1 + n - 1) (\ell_2 + n - 1) \cdots (\ell_r + n - 1)
	\end{equation}
so that
	\begin{equation}
	F(n)! = n!  \prod_{i = 1}^{r} \frac{(\ell_i + n - 1)!}{\ell_i^n (\ell_i - 1)!} =
	\Gamma(n+1) \prod_{i = 1}^{r} \frac{\Gamma(\ell_i+n)}{\ell_i^n \Gamma(\ell_i)}. 
	\label{F(n)factorial1}
	\end{equation}
Consequently, Eq.~(\ref{norme1s}) can be written as 
	\begin{equation}
	\vert{\cal N}\vert^{2} = \sum_{n=0}^{s-1} \frac{1}{n!} 
	\frac{(\ell_1)_n}{\ell_1^n} 
	\frac{(\ell_2)_n}{\ell_2^n} \cdots
	\frac{(\ell_r)_n}{\ell_r^n}
	\vert z \vert^{2n}
	\label{77}
	\end{equation}
where $(a)_n = \Gamma(a+n) / \Gamma(a)$ is the Pochhammer symbol. A long calculation, using the inverse Mellin transform \cite{Bateman}, shows that the states given by (\ref{KP-CS-truncated}) and (\ref{77}) 
satisfy the over-completeness relation 
	\begin{equation}
	\int d\mu (\vert z \vert) \vert z , \varphi \rangle \langle z , \varphi \vert = \sum_{n=0}^{s-1} \vert n \rangle \langle n \vert
	\end{equation}
with the measure 
	\begin{equation}
	d \mu (|z|) = \frac{1}{\pi} M(|z|^2) \vert {\cal N} \vert^{2} d^2 z
	\label{mesure-alg-s}
	\end{equation}
where  
	\begin{equation}
	M(\vert z \vert^2) = \frac{\Gamma(\ell_1)\Gamma(\ell_2)\cdots
	\Gamma(\ell_r)}{\ell_1 \ell_2\cdots \ell_r} G_{r,1}^{1,0}\bigg( \frac{\vert z
	\vert^2}{\ell_1\ell_2 \cdots \ell_r} \bigg\vert {}_{~~~~~~~~~0}^{\ell_1 - 1, \ell_2
	- 1, \cdots, \ell_r - 1} \bigg) 
	\label{mesure1}
	\end{equation}
$G$ being the Meijer function defined in \cite{Erdelyi, Kobayashi, Prudnikov}.

For $r=1$ and $\kappa_1 = 0$ (corresponding to the limiting case $\ell_1 \to \infty$), Eqs.~(\ref{KP-CS-truncated}) and 
(\ref{norme1s}) give the coherent states
	\begin{equation}
	\vert z , \varphi \rangle = \bigg( \sum_{n=0}^{s-1} \frac{1}{n!} \vert z \vert^{2n}\bigg)^{-1} 
	\sum_{n=0}^{s-1} \frac{1}{\sqrt{n!}} (z e^{-i \varphi})^n \vert n \rangle
	\end{equation}
which coincide with the coherent states for the Pegg and Barnett oscillator discussed in \cite{0411210}.

%%%%%%%%%%%%%%%%%%%%%%%%%%%%%%%%%%%%%%%%%%%%%%%%%%%%%%%%%%%%%%%%%%%%%
\subsection{Finite-dimensional Fock-Hilbert space}
%%%%%%%%%%%%%%%%%%%%%%%%%%%%%%%%%%%%%%%%%%%%%%%%%%%%%%%%%%%%%%%%%%%%%

The search for coherent states 
	\begin{equation}
	\vert {{z}} , \varphi \rangle = \sum_{n=0}^{d-1} {\overline{b_n}} {{z}}^n \vert n \rangle
	\end{equation}
in the situation where 
	\begin{equation}
	\kappa_1 = - \frac{1}{d-1} < 0 \qquad \kappa_i \geq 0 \quad (i=2, 3, \cdots, r)
	\label{kappa cas fini}
	\end{equation}
can be done on the same pattern as in Sections 5.1 and 5.2 by starting from the correspondence
	\begin{equation}
	a^- \longrightarrow \frac{d}{d{{z}}} \qquad 
	|n \rangle \longrightarrow b_{n} {{z}}^{n} \equiv \langle \bar {{z}} , \varphi \vert n \rangle
	\end{equation}
which is compatible with 
	\begin{equation}
	N \longrightarrow {{z}} \frac{d}{d{{z}}} \qquad 
	a^+ \longrightarrow {{z}} \bigg( 1 - \frac{1}{d-1} {{z}}
	\frac{d}{d{{z}}}\bigg) \bigg( 1 + \kappa_2 {{z}}
	\frac{d}{d{{z}}}\bigg)\cdots \bigg( 1 + \kappa_r {{z}}
	\frac{d}{d{{z}}}\bigg).
	\label{corres pr N et aplus pr d}
	\end{equation}
We thus get 
	\begin{equation}
	\vert {{z}} , \varphi \rangle = {\cal N}^{-1}
	\sum_{n=0}^{d-1} \frac{\sqrt{F(n)!}}{n!} {{z}}^n e^{-i F(n)\varphi} \vert n \rangle \qquad 
	\vert {\cal N} \vert^{2} = \sum_{n=0}^{d-1} \frac{F(n)!}{(n!)^2} \vert {{z}} \vert^{2n}
	\label{bidule}
	\end{equation}
where $F(n)!$ follows from (\ref{F(n) factorial}) and (\ref{Fn fini}). The 
$\vert {{z}} , \varphi \rangle$ states are temporally stable. They satisfy 
	\begin{equation}
	\vert {{z}} , \varphi \rangle = {\cal N}^{-1} \exp( {{z}}  a^+)  \vert 0 \rangle
	\end{equation}
and are thus coherent states in the Perelomov sense. 

In the case 
	\begin{equation}
	\ell_i = \frac{1}{\kappa_i} \in \mathbb{N}^* \quad (i = 2,3, \cdots, r)
	\end{equation}
the $F(n)!$ generalized factorial in (\ref{bidule}) can be calculated to be 
	\begin{equation}
	F(n)! =  n! \frac{(d-1)!}{(d-1)^n (d - 1 - n)!} \prod_{i = 2}^{r}
	\frac{(\ell_i + n - 1)!}{\ell_i^n (\ell_i - 1)!}. 
	\label{F(n) factoriel cas fini}
	\end{equation}
It can be shown, using the inverse Mellin transform \cite{Bateman}, that the coherent states given by (\ref{bidule}) and (\ref{F(n) factoriel cas fini}) 
satisfy the over-completeness relation 
	\begin{equation}
	\int d\mu(\vert {{z}} \vert) \vert  {{z}}, \varphi \rangle \langle  {{z}}, \varphi \vert = \sum_{n=0}^{d-1} \vert n \rangle \langle n \vert
	\end{equation}
where the $d\mu$ measure reads 
	\begin{equation}
	d \mu (\vert {{z}} \vert) = \frac{1}{\pi} M(\vert {{z}} \vert^2) \vert{ \cal N }\vert^{2} d^2 {{z}}
	\end{equation}
where
	\begin{equation}
	M (\vert {{z}} \vert^2) = \frac{1}{(d-1)\Gamma(d)}
	\frac{\Gamma(\ell_2)\Gamma(\ell_3)\cdots \Gamma(\ell_r)}{\ell_2 \ell_3 \cdots \ell_r}
	G_{r,1}^{1,1}\bigg( \frac{\vert {{z}} \vert^2}{(d-1)\ell_2\cdots \ell_r}
	\bigg\vert {}_{~~~~~~~~~0 }^{-d, \ell_2 - 1, \cdots, \ell_r - 1} \bigg)
	\end{equation}
in term of the $G$ Meijer function.

As an example, let us consider the $r=1$ case. From (\ref{bidule}) and (\ref{F(n) factoriel cas fini}), we 
have
	\begin{equation}
	\vert {{z}} , \varphi \rangle = {\cal N}^{-1} \sum_{n = 0}^{d-1} 
	\sqrt{ \frac{1}{n!} \frac{(d-1)!}{(d-1)^n (d-1-n)!} } {{z}}^n
	e^{-iF(n)\varphi}\vert n \rangle
	\end{equation}
where the normalization factor follows from 
	\begin{equation}
	\vert{\cal N}\vert^{2} = \bigg (1 +
	\frac{\vert {{z}} \vert^2}{d-1} \bigg)^{d-1}.
	\end{equation}
In addition, the $d\mu$ measure is given here by
	\begin{equation}
	d \mu (\vert {{z}} \vert) = \frac{1}{\pi} \frac{d}{d-1} \bigg(1 +  \frac{\vert {{z}} \vert^2}{d-1} \bigg)^{-2}
	\end{equation}
(the expression for $G_{1,1}^{1,1}$ is taken from \cite{Kobayashi}).

%%%%%%%%%%%%%%%%%%%%%%%%%%%%%%%%%%%%%%%%%%%%%%%%%%%%%%%%%%%%%%%%%%%%%%%%%%%%%%%%%%%%%%%
\section{{\it \`a la} Barut--Girardello coherent states}
%%%%%%%%%%%%%%%%%%%%%%%%%%%%%%%%%%%%%%%%%%%%%%%%%%%%%%%%%%%%%%%%%%%%%%%%%%%%%%%%%%%%%%%%%

%%%%%%%%%%%%%%%%%%%%%%%%%%%%%%%%%%%%%%%%%%%%%%%%%%%%%%%%%%%
\subsection{ Infinite-dimensional Fock-Hilbert space}
%%%%%%%%%%%%%%%%%%%%%%%%%%%%%%%%%%%%%%%%%%%%%%%%%%%%%%%%%%%

Going back to $\kappa_i \geq 0$ ($i = 1,2, \cdots, r$), we now look for coherent states 
	\begin{equation}
	\vert {{z}} , \varphi \rangle =  \sum_{n = 0}^{\infty} {\overline{c_n}} {{z}}^n \vert n \rangle
	\label{71}
	\end{equation}
in a realization of ${\cal A}_{\{ \kappa \}}$ in term of complex variable ${{z}}$ with 
	\begin{equation}
	\vert n \rangle \longrightarrow c_{n} {{z}}^{n} \equiv \langle \bar {{z}} , \varphi \vert n \rangle \qquad
	a^+ \longrightarrow {{z}}
	\label{correspondance1BG}
	\end{equation}
($a^+$acts as a simple multiplication by ${{z}}$). The combination of (\ref{action sur les n}) with (\ref{correspondance1BG}) 
leads to 
	\begin{equation}
	c_{n} = \sqrt{F(n+1)} e^{-i[F(n+1) - F(n)] \varphi} c_{n+1}
	\label{recursion cn}
	\end{equation}
which can be iterated to give
	\begin{equation}
	c_{n} = c_0 \frac{1}{\sqrt{F(n)!}} e^{+iF(n)\varphi}.
	\end{equation}
We are thus left with states
	\begin{equation}
	\vert {{z}} , \varphi \rangle = {\cal N}^{-1}
	\sum_{n=0}^{\infty} \frac{1}{\sqrt{F(n)!}} {{z}}^n e^{-iF(n)\varphi} \vert n \rangle 
	\label{74}
	\end{equation}
with a normalization factor given by 
	\begin{equation}
	\vert{\cal N}\vert^{2} = \sum_{n=0}^{\infty} \frac{1}{F(n)!} \vert {{z}} \vert^{2n}
	\label{75}
	\end{equation}
where $F(n)!$ follows from (\ref{F(n) factorial}) and (\ref{Fn infini}). The series in (\ref{74}) and (\ref{75})
converges in the whole complex plane $\mathbb{C}$. Note that the correspondence (\ref{correspondance1BG}) can be 
completed by 
	\begin{equation}
	N \longrightarrow {{z}} \frac{d}{d{{z}}} \qquad
	a^- \longrightarrow  \bigg( 1 + \kappa_1 {{z}}
	\frac{d}{d{{z}}}\bigg) \bigg( 1 + \kappa_2 {{z}}
	\frac{d}{d{{z}}}\bigg) \cdots \bigg( 1 + \kappa_r {{z}}
	\frac{d}{d{{z}}}\bigg) \frac{d}{d{{z}}}. 
	\end{equation}
cf.~(\ref{corres pr N et aplus}) and (\ref{corres pr N et aplus pr d}).

The $\vert {{z}} , \varphi \rangle$ vectors 
are eigenstates of the $a^-$ annihilation operator 
	\begin{equation}
	a^- \vert {{z}} , \varphi \rangle = {{z}} \vert {{z}} , \varphi
	\rangle \label{eigenvalues-amoins}
	\end{equation}
and can thus be called coherent states in the Barut--Girardello sense \cite{Barut}. Furthermore 
	\begin{equation}
	e^{-i H t} \vert {{z}} , \varphi \rangle = \vert {{z}} , \varphi + t \rangle
	\end{equation}
which means that they are stable under time evolution. 

Let us continue with the special case where each $\ell_i = 1 / \kappa_i$ ($i= 1, 2, \cdots, r$) is a strictly positive integer 
like in (\ref{un sur kappai entier}). In this case, Eq.~(\ref{75}) can be written as  
	\begin{equation}
	\vert{\cal N}\vert^{2} = \sum_{n=0}^{\infty} \frac{1}{n!} 
	\frac{\ell_1^n}{(\ell_1)_n} 
	\frac{\ell_2^n}{(\ell_2)_n}
	\cdots 
	\frac{\ell_r^n}{(\ell_r)_n}
	\vert {{z}} \vert^{2n}
	\label{correction}
	\end{equation}
or alternatively 
	\begin{equation}
	\vert{\cal N}\vert^{2} = {}_0F_r(\ell_1,\ell_2, \cdots, \ell_r; \ell_1\ell_2\cdots \ell_r\vert {{z}} \vert^2)
	\end{equation}
in term of the generalized hypergeometric function. We can show that the $\vert {{z}} , \varphi \rangle$ 
coherent states given by Eqs.~(\ref{74}) and (\ref{correction}) satisfy the over-completeness relation 
	\begin{equation}
	\int d\mu (\vert {{z}} \vert) \vert {{z}} , \varphi \rangle \langle {{z}} , \varphi \vert = \sum_{n=0}^{\infty} \vert n \rangle \langle n \vert
	\end{equation}
with the measure
	\begin{equation}
	d\mu (\vert {{z}} \vert) = \frac{1}{\pi} M(\vert {{z}} \vert^2) \vert {\cal N} \vert^2 d^2 {{z}} 
	\label{measureBG}
	\end{equation}
where 
	\begin{equation}
	M(\vert {{z}} \vert^2) = \frac{\ell_1 \ell_2\cdots \ell_r}{\Gamma(\ell_1)\Gamma(\ell_2)\cdots \Gamma(\ell_r)} 
	G_{0,r+1}^{r+1,0}\bigg(\ell_1 \ell_2\cdots \ell_r| {{z}}|^2 \bigg\vert 0, \ell_1 - 1, \ell_2 - 1, \cdots, \ell_r - 1\bigg)
	\label{MdeBG}
	\end{equation}
in term of the $G$ Meijer function. 

Finally, it is interesting to examine the $r = 1$ particular case 
corresponding to algebra ${\cal A}_{\kappa}$ with $\kappa = 1/\ell$, $\ell \in \mathbb{N}^*$. Then, 
the coherent states are
	\begin{equation}
	\vert {{z}}, \varphi \rangle = {\cal N}^{-1} \sum_{n=0}^{\infty} 
	\sqrt{ \frac{1}{n!} \frac{\ell^n(\ell-1)!}{(\ell + n -1)!} } {{z}}^{n} e^{-iF(n)\varphi} |n \rangle
	\label{stateGK}
	\end{equation}
where
	\begin{equation}
	\vert{\cal N}\vert^2 = \sum_{n=0}^{\infty}
	\frac{1}{n!} \frac{\ell^n(\ell-1)!}{(\ell + n - 1)!} |{{z}}|^{2n} = {}_0F_1(\ell ; \ell \vert {{z}} \vert^2).
	\label{normGK}
	\end{equation}
Taking the Meijer function from \cite{Kobayashi}, we have
	\begin{equation}
	d \mu (\vert {{z}} \vert) = \frac{1}{\pi} 2\ell K_{\ell-1}(2\sqrt{\ell}\vert {{z}} \vert) 
	                                                I_{\ell-1}(2\sqrt{\ell}\vert {{z}} \vert) d^2{{z}}
	\label{9292}
	\end{equation}
where $I$ and $K$ are modified Bessel functions of first and second kind, 
respectively. 

At this point, a contact with some previous works is in order. First, the states given 
by (\ref{74}) and (\ref{correction}) can be viewed as Gazeau--Klauder \cite{Gazeau1, Klauder2} 
coherent states associated with a quantum mechanical system whose Hamiltonian is 
$F(N)$. Second, Eqs.~(\ref{stateGK})-(\ref{9292}) corresponds to Gazeau-Klauder 
coherent states for quadratic (in $N$) Hamiltonians like those for the infinite well and 
P\"oschl-Teller systems \cite{Antoine, Gazeau1}. 

%%%%%%%%%%%%%%%%%%%%%%%%%%%%%%%%%%%%%%%%%%%%%%%%%%%%%%%%%%%%%%%%%%%%%
\subsection{Finite-dimensional Fock-Hilbert space}
%%%%%%%%%%%%%%%%%%%%%%%%%%%%%%%%%%%%%%%%%%%%%%%%%%%%%%%%%%%%%%%%%%%%%

\subsubsection{Preliminary observation}

It is natural to ask if the realization of ${\cal A}_{ \{ \kappa \} }$ of 
Section 6.1, in which creation operator $a^+$ acts as multiplication, works 
in the finite-dimensional case where $n = 0, 1, \cdots, d-1$. The recursion 
relation (\ref{recursion cn}) is valid for $n = 0, 1, \cdots, d-2$. For $n = d-1$, 
the action of $a^+$ on the extremal state $\vert d - 1 \rangle$ is zero. Thus, 
according to (\ref{correspondance1BG}), coefficient $c_{d-1}$ satisfies 
	\begin{equation}
	c_{d-1} {{z}}^d = 0.
	\label{eq99}
	\end{equation}
Equation (\ref{eq99}) admits the solution $c_{d-1} = 0$ and from (\ref{recursion cn}) 
it follows that $c_{d-1} = c_{d-2} = \cdots = c_0 = 0$ leading to trivial and 
thus unacceptable coherent states. Another solution of (\ref{eq99}) is  
	\begin{equation}
	{{z}}^d = 0
	\label{gragra}
	\end{equation}
which is reminiscent of the $k$-fermions (with $k = d$) discussed in Section 3 and 
which shows that ${{z}}$ cannot be here a complex variable but should be considered 
as a generalized Grassmann variable of order $d$. As a conclusion, the construction of 
coherent states {\it \`a la} Barut--Girardello for algebra ${\cal A}_{ \{ \kappa \} }$ 
with a finite-dimensional representation requires the introduction of Grassmann 
variables. This point is be developed in the remaining part of this section for the 
${\cal A}_{\{ \kappa \}}$ algebra with $\kappa_1 < 0$ and the ${\cal A}_{\{ \kappa \},s}$ 
truncated algebra with $\kappa_1 \geq 0$.

\subsubsection{$d$-fermionic coherent state for ${\cal A}_{\{ \kappa \}}$}

This section concerns the situation where $\kappa_1 < 0$ and $\kappa_i \geq 0$ ($i = 2, 3, \cdots, r$) 
for which $F(n)$ is given by (\ref{Fn fini}). Guided by (\ref{gragra}) with the ${{z}}$ complex variable 
replaced by a $\theta$ generalized Grassmann variable and applying the same {\it \`a la} Fock--Bargmann approach 
as in Section 6.1, we obtain the following unnormalized states 
	\begin{eqnarray}
	\vert \theta , \varphi \rangle = \sum_{n=0}^{d-1} 
	\frac{1}{\sqrt{F(n)!}} \theta^n e^{-i F(n) \varphi} \vert n \rangle
	\label{grassman BG}
	\end{eqnarray}
where $F(n)!$ follows from (\ref{F(n) factorial}) and (\ref{Fn fini}). The normalization follows from 
	\begin{eqnarray}
	\langle \theta , \varphi \vert \theta , \varphi \rangle = 
	\sum_{n=0}^{d-1} \exp \bigg[ - i \pi \frac{n(n-1)}{2d} \bigg] \frac{1}{F(n)!} (\bar\theta \theta)^n
	\label{norm of grassman BG}
	\end{eqnarray}
where Eq.~(\ref{eq56}) is taken into account. 

In view of $\theta^d = 0$, the $\vert \theta , \varphi \rangle$ states can be called $d$-fermionic coherent states \cite{DHKrusse}. They 
satisfy
    \begin{eqnarray}
		a^- \vert \theta, \varphi \rangle = \theta \vert \theta, \varphi \rangle
    \end{eqnarray}
and are thus coherent states in the Barut--Girardello sense. In addition, they are stable under time evolution, i.e. 
    \begin{eqnarray}
		e^{-i H t}\vert \theta , \varphi \rangle = \vert \theta , \varphi + t \rangle.
    \end{eqnarray}
Finally, states (\ref{grassman BG}) constitute an over-complete set with
	\begin{equation}
	\int \vert \theta , \varphi \rangle d\mu (\theta ,\bar \theta )
	\langle \theta , \varphi \vert = \sum_{n = 0}^{d-1} \vert n \rangle \langle n \vert
	\end{equation}
for the $d \mu$ measure satisfying the following  integral formula
	\begin{equation}
	\frac{1}{F(n)!} \int \theta ^n \; d\mu (\theta ,\bar \theta ) \; \bar \theta^m = \delta_{n,m}. 
	\label{condition mesure}
	\end{equation}
It can be proved that $d \mu$ is given by 
	\begin{equation}
	d\mu (\theta ,\bar \theta ) = \sum_{n = 0}^{d-1} F(n)! ~ d\theta ~ \theta^{d-1-n} ~ \bar\theta^{d-1-n} ~ d\bar\theta
	\label{mesure grassmann}
	\end{equation}
owing to integration rules (\ref{majid-formule1}) and (\ref{majid-formule2}).

As an example, let us take $r=1$ corresponding to ${\cal A}_{\kappa}$ with 
$\kappa \equiv \kappa_1 < 0$ and $-1 / \kappa \in \mathbb{N}^*$. The latter 
algebra can be identified to $su(2)$ with generators $J_-$, $J_+$ and $J_3$ \cite{Atakishiyev}
\begin{eqnarray}
J_- = \frac{1}{\sqrt{-\kappa}} a^-          \qquad 
J_+ = \frac{1}{\sqrt{-\kappa}} a^+          \qquad 
J_3 = \frac{1}{2 \kappa} (I + 2 \kappa N). 
\label{generators SU(2)}
\end{eqnarray}
By making the substitutions 
	\begin{eqnarray}
	\vert n \rangle \longleftrightarrow \vert j , m \rangle  \qquad  
	n \longleftrightarrow  j + m \qquad
	-\frac{1}{\kappa} \longleftrightarrow 2j \qquad
	d \longleftrightarrow 2j + 1 
	\end{eqnarray}
where $m = - j, -j+1, \cdots, +j$ in terms of angular momentum or $su(2)$ quantum numbers, 
Eq.~(\ref{grassman BG}) can be specialized to 
	\begin{eqnarray}
	\vert \theta , \varphi \rangle = \sum_{m=-j}^{+j}
	\sqrt{ \frac{1}{(2j)!} \frac{(j-m)!}{(j+m)!} } \bigg( \sqrt{2j} \theta \bigg)^{j+m}
	e^{-i F(j+m) \varphi} \vert j ,m \rangle. 
	\label{su2 grassman BG}
	\end{eqnarray}
These states satisfy the eigenvalue equation
	\begin{eqnarray}
J_- |\theta , \varphi \rangle = \sqrt{2j} \theta |\theta , \varphi \rangle
	\end{eqnarray}
and thus can be viewed as $su(2)$ Barut--Girardello coherent states labeled by a 
$\theta$ generalized Grassmann variable of order $2j+1$ ($\theta^{2j+1} = 0$).

Two extremal cases are of interest. First for $d=2$, Eq.~(\ref{su2 grassman BG}) 
gives the states 
	\begin{eqnarray}
	\vert \theta , \varphi \rangle = \vert 1/2 , - 1/2 \rangle +
	           \theta e^{-i \varphi} \vert 1/2 , + 1/2 \rangle = 
 \vert 0 \rangle + 
	           \theta e^{-i \varphi} \vert 1 \rangle 
	\label{su2 grassman BG pr d=2}
	\end{eqnarray}
which for $\varphi = 0$ coincide with the coherent states for the fermionic oscillator \cite{Berube} 
or a qubit. Second in the limiting case $d \to \infty$ (or $\kappa \to 0^-$), we can identify $\theta$ 
to a complex variable, say $z$, and (\ref{grassman BG}) leads to the states 
	\begin{eqnarray}
	\vert z , \varphi \rangle = \sum_{n=0}^{\infty}
	\sqrt{ \frac{1}{n!} } z^n e^{-i n \varphi} \vert n \rangle
	\end{eqnarray}
which look like the Glauber coherent states for the bosonic oscillator up to a re-labeling of $z \exp(-i \varphi)$ as $z$. 

\subsubsection{$s$-fermionic coherent state for ${\cal A}_{\{ \kappa \},s}$}

For the ${\cal A}_{\{ \kappa \},s}$ truncated algebra with $\kappa_i \geq 0$ ($i = 1, 2, \cdots, r$), the approach developed 
in 6.2.2 for deriving Barut--Girardello coherent states is valid under the condition to replace $d$ by $s$, $a^-$ by $a^-(s)$ 
and $F(n)$ calculated via (\ref{Fn fini}) by $F(n)$ calculated via (\ref{Fn infini}). This leads to coherent states of type 
(\ref{grassman BG}) with $d=s$. 

As an example, let us consider the case where $r=1$ and $\kappa \equiv \kappa_1 >0$. The corresponding algebra, 
${\cal A}_{\kappa}$ can be identified with $su(1,1)$ with generators $K_-$, $K_+$ and $K_3$ given by \cite{Atakishiyev}
      \begin{eqnarray}
K_- = \frac{1}{\sqrt{\kappa}} a^-           \qquad 
K_+ = \frac{1}{\sqrt{\kappa}} a^+           \qquad 
K_3 = \frac{1}{2 \kappa} (I + 2 \kappa N). 
      \label{generators SU(1,1)}
      \end{eqnarray}
To pass from the representation of ${\cal A}_{\kappa}$ to that of $su(1,1)$ we make the substitutions       
      \begin{eqnarray}      
\vert n \rangle \longleftrightarrow \vert b , b + n \rangle \qquad
n \longleftrightarrow n \qquad
\frac{1}{\kappa} \longleftrightarrow 2b
      \end{eqnarray}
where $b$ is the Bargmann index (generally noted $k$ but here noted $b$ in order to avoid confusion with $k$ in 
$k$-fermions). The truncation to order $s$ of the infinite-dimensional discrete representation of $su(1,1)$ 
yields the coherent states 
	\begin{eqnarray}
	\vert \theta , \varphi \rangle = \sum_{n=0}^{s-1}
  \sqrt{ \frac{(2b-1)!}{n!(2b + n - 1)!} } \bigg( \sqrt{2b} \theta \bigg)^{n}
	e^{-i F(n) \varphi} \vert b , b + n \rangle
	\label{su11 BG troncated}
	\end{eqnarray}
which satisfy 
	\begin{eqnarray}
K_-(s) |\theta , \varphi \rangle = \sqrt{2b} \theta  |\theta , \varphi \rangle
	\end{eqnarray}
where $K_-(s) = \sqrt{2b} a^-(s)$. 

We close this section with the truncated oscillator algebra. This algebra 
is described by Eqs.~(\ref{operators tronques aplus})-(\ref{action des b sur les n bmoins et N}) with 
$F(n) = n$ ($\kappa_i = 0$ for $i = 1, 2, \cdots, r$). The corresponding $s$-fermionic coherent states are simply
 	\begin{eqnarray}
	\vert \theta , \varphi \rangle = \sum_{n=0}^{s-1}
	\frac{1}{\sqrt{n!}} {\theta}^{n} e^{-i n \varphi} \vert n \rangle.
	\end{eqnarray}
They satisfy
 	\begin{eqnarray}
a^-(s) |\theta , \varphi \rangle  = \theta |\theta , \varphi \rangle
	\end{eqnarray}
and can be refered to Barut--Girardello coherent states for the Pegg--Barnett oscillator. 

%%%%%%%%%%%%%%%%%%%%%%%%%%%%%%%%%%%%%%%%%%%%%%%%%%%%%%%%%%%%%%%
\section{Closing remarks}
%%%%%%%%%%%%%%%%%%%%%%%%%%%%%%%%%%%%%%%%%%%%%%%%%%%%%%%%%%%%%%%

The ${\cal A}_{ \{ \kappa \} }$ polynomial Weyl-Heisenberg algebra considered in this paper generalizes, along the 
line of the works in \cite{Carballo, newdaokib, daoud-kibler1, daoud-kibler2, Daskaloyannis, Fernandez, Gavrilik, 
Katriel, Man'ko, Quesne}, the ${\cal A}_{\kappa}$ algebra introduced in \cite{daoud-kibler2}. Indeed, the 
${\cal A}_{ \{ \kappa \} }$ algebra, characterized by structure function $F$ such that $F(N)$ is a polynomial of 
order $r+1$ in number operator $N$, gives for $r=1$ the ${\cal A}_{ \kappa }$ algebra which gives in turn 
the $su(2)$, $su(1,1)$ and oscillator algebras as very special cases. Note that the ${\cal A}_{ \{ \kappa \} }$ 
algebra can be viewed as a special class of the $f$-oscillators introduced by Man'ko {\it et al} 
\cite{Man'ko}. The dimension of the Fock-Hilbert representation space of ${\cal A}_{ \{ \kappa \} }$, with 
$\{ \kappa \} \equiv \{ \kappa_1, \kappa_2, \cdots, \kappa_r \}$, depends on the sign of the 
$\kappa_i$ parameters ($i = 1, 2, \cdots, r$). In the case of an infinite-dimensional space, 
we used a truncation procedure of the Pegg--Barnett type for generating a ${\cal A}_{ \{ \kappa \} , s }$ 
truncated algebra of truncation order $s$. As a result, of importance for deriving 
certain coherent states, a connection was established between the $s$-fermionic algebra 
(arising from two quon algebras) and either the ${\cal A}_{ \{ \kappa \} , s }$ truncated 
algebra or the ${\cal A}_{ \{ \kappa \} }$ algebra with a representation of finite dimension 
($d=s$). Note that such a connection can be derived between any generalized Weyl-Heisenberg 
algebra with a representation of dimension $d=s$ and the $s$-fermionic algebra. 

Very few papers were devoted to the construction of coherent states of the Perelomov type (i.e., resulting from 
the action of a displacement operator on the vacuum) for generalized Weyl-Heisenberg algebras in finite or infinite 
dimension. We may mention Ref.~\cite{El Kinani1} corresponding to a polynomial Weyl-Heisenberg algebra with $r=1$. The 
approach developed in \cite{El Kinani1} is very difficult to adapt to the case $r \geq 2$. In the present paper, Perelomov 
coherent states were worked out in a Fock--Bargmann realization where the annihilation operator of ${\cal A}_{ \{ \kappa \} }$ 
and ${\cal A}_{ \{ \kappa \} , s }$ acts as derivative by a bosonic complex variable. For a infinite-dimensional 
representation of ${\cal A}_{ \{ \kappa \} }$, it was shown that bosonic Peremolov coherent states exist 
if and only if $r=1$. For a finite-dimensional representation of ${\cal A}_{ \{ \kappa \} }$ or 
${\cal A}_{ \{ \kappa \} , s }$, bosonic Perelomov coherent states were constructed for arbitrary $r$; to the best 
of our knowledge, this constitutes the first derivation of Peremolov coherent states in finite dimension.

The situation is different for coherent states of the Barut--Girardello type. Most of the works on Barut--Girardello 
coherent states were achieved for infinite-dimensional representations of generalized Weyl-Heisenberg algebras by 
diagonalizing an annihilation operator (for example, see \cite{Fernandez, Man'ko}). In this paper, bosonic 
Barut--Girardello coherent states for an infinite-dimensional representation of ${\cal A}_{ \{ \kappa \} }$
were derived in a Fock--Bargmann realization where the creation operator acts as multiplication  
by a bosonic complex variable. For finite-dimensional representation of ${\cal A}_{ \{ \kappa \} }$ and 
${\cal A}_{ \{ \kappa \} , s }$, the construction of Barut--Girardello coherent states required the introduction 
of generalized Grassmann variables (with the creation operator acting as multiplication by a 
$k$-fermionic or generalized Grassmann variable). The construction of Barut--Girardello coherent states in finite 
dimension (illustrated with the $su(2)$ algebra, the $su(1,1)$ truncated algebra and the Pegg--Barnett oscillator 
algebra) is realized in this article for the first time. 

The (bosonic) Perelomov coherent states as well as the (bosonic and $k$-fermionic) Barut-Girardello coherent states 
constructed in this work satisfy continuity, temporal stability under time evolution and an over-completeness 
property. For bosonic Perelomov and Barut--Girardello coherent states, the over-completeness property was derived 
by means of a measure using the $G$ Meijer function and for $k$-fermionic Barut--Girardello coherent states the 
relevant measure was obtained by using the Majid--Rodr\'\i guez-Plaza integration formula for generalized Grassmann 
variables.   

As a possible extension of this work, let us mention the derivation of those states of ${\cal A}_{ \{ \kappa \} }$ which 
minimize the Robertston-Schr\"odinger \cite{Shrodinger1, Robertson} uncertainty relation. It would be interesting to see 
how these states (called intelligent states in the cases of $su(2)$ and $su(1,1)$ \cite{Aragone1, Trifonov2, Trifonov}) 
incorporate in a unified scheme Perelomov and Barut--Girardello states. Moreover, another interesting extension would 
be to construct coherent state vectors, in the sense of Ali--Engli$\check{{\rm s}}$--Gazeau \cite{AEG}, for the 
${\cal A}_{ \{ \kappa \} }$ and ${\cal A}_{ \{ \kappa \} , s }$ algebras. 

It should be emphasized that most of the generalized Weyl-Heisenberg algebras considered in the 
literature have infinite-dimensional representations. The consideration in the present 
paper of generalized Weyl-Heisenberg algebras with finite-dimensional representations, along the line 
initiated in \cite{daoud-kibler2, daoud-kibler3}, opens an avenue of future developments, especially 
in quantum information and quantum computation where finite-dimensional Hilbert spaces are of 
paramount importance. We hope that the results contained in this paper (in particular, the Barut--Girardello 
coherent states in finite dimension) would be useful in quantum information where intrication of coherent 
states plays a fundamental role. As encouraging preliminary works, the ${\cal A}_{\kappa}$ algebra already 
proved to be useful \cite{daoud-kibler2, daoud-kibler3}, for deriving mutually unbiased bases of interest in 
quantum cryptography and quantum state tomography, and intrication of $k$-fermionic coherent states (defined in 
\cite{DHKrusse}) was studied in \cite{Maleki}.

\section*{Acknowledgments} 

MD would like to thank the hospitality and kindness of the 
{\em Groupe de physique th\'eorique de l'Institut de 
Physique Nucl\'eaire de Lyon} where this work was done. 

%%%%%%%%%%%%%%%%%%%%%%%%%%%%%%%%%%%%%%%%%%%%%%%%%%%%%%%%%%%%%%%%%%%%%%
%%%%%%%%%%%%%%%%%%%%%%%%%%%%%%%%%%%%%%%%%%%%%%%%%%%%%%%%%%%%%%%%%%%%%%

\end{document}